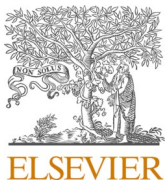
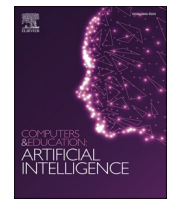

# Evaluating adaptive and generative AI-based feedback and recommendations in a knowledge-graph-integrated programming learning system

Lalita Na Nongkhai [a], Jingyun Wang [b,*], Adam Wynn [b], Takahiko Mendori [a]

[a] *Graduate School of Engineering, Kochi University of Technology, Kochi, Japan*
[b] *Department of Computer Science, Durham University, Durham, UK*

ABSTRACT

This paper introduces the design and development of a framework that integrates a large language model (LLM) with a retrieval-augmented generation (RAG) approach leveraging both a knowledge graph and user interaction history. The framework is incorporated into a previously developed adaptive learning support system to assess learners' code, generate formative feedback, and recommend exercises. Moreover, this study examines learner preferences across three instructional modes: adaptive, Generative AI (GenAI), and hybrid GenAI–adaptive. An experimental study was conducted to compare the learning performance and perception of the learners, and the effectiveness of these three modes using four key log features derived from 4956 code submissions across all experimental groups. The analysis results show that learners receiving feedback from GenAI modes had significantly more correct code and fewer code submissions missing essential programming logic than those receiving feedback from adaptive mode. In particular, the hybrid GenAI–adaptive mode achieved the highest number of correct submissions and the fewest incorrect or incomplete attempts, outperforming both the adaptive-only and GenAI-only modes. Questionnaire responses further indicated that GenAI-generated feedback was widely perceived as helpful, while all modes were rated positively for ease of use and usefulness. These results suggest that the hybrid GenAI–adaptive mode outperforms the other two modes across all measured log features.

## 1. Introduction

The education of computer programming has long been challenging. Novice learners frequently encounter problems in understanding programming concepts, constructing logical solutions, and translating these solutions into code (Derus & Ali, 2012). To address these challenges, the recommendation of personalized learning pathways and the provision of tailored content based on individual learners' performance have been proposed (Chookaew et al., 2014; Gavrilović et al., 2018). While consistent engagement in independent problem-solving tasks has been proven to be effective for the enhancement of programming skills (Djambong & Freiman, 2016), numerous studies have proposed to present adaptive programming exercises in response to learners' performance or cognitive levels (Fabic et al., 2018; Zheng et al., 2022).

In addition to personalized programming practice, timely formative feedback on programming code is also essential. As defined by Shute (2008), feedback can be understood as information provided to learners with the aim of guiding them to reflect on learning, helping them eliminate misconceptions, or reinforce positive behaviors. Moreover, Garcia (2025) highlighted the importance of timely feedback as a key element in scaffolding effective learning, particularly within programming education. However, when learners practice programming independently online, they often lack formative feedback and receive only correctness results or error messages.

To provide detailed feedback, numerous researchers have explored the adoption of various artificial intelligence (AI) techniques (Deeva et al., 2021; Frankford et al., 2025). More recently, large language models (LLMs), a specialized subset of machine learning models optimized for language tasks, have experienced rapid development and widespread adoption, particularly in the field of Generative AI (GenAI), leading to growing interest in their application for programming education. Emerging studies have explored the use of LLMs to help with assignments and generate code explanations, provide debugging assistance, and deliver personalized formative feedback (Kosar et al., 2024; George & Dewan, 2024; Majumdar et al., 2024). Additionally, retrieval-augmented generation (RAG) techniques have been introduced to enhance the relevance and accuracy of AI-generated feedback by integrating contextual or domain-specific knowledge, such as code repositories or knowledge graphs (Li et al., 2024b). Together, these approaches highlight the potential of GenAI to support personalized programming practice.

* Corresponding author.
*E-mail addresses:* lalita.nanongkhai88@gmail.com (L. Na Nongkhai), jingyun.wang@durham.ac.uk (J. Wang), adam.t.wynn@durham.ac.uk (A. Wynn), mendori.takahiko@kochi-tech.ac.jp (T. Mendori).

In our previous work (Na Nongkhai et al., 2023, 2025), we introduced ADVENTURE (ADaptiVe lEarning support system based on oNTology of mUltiple pRogramming languagEs), an adaptive learning support system designed to facilitate programming exercises across multiple languages. ADVENTURE incorporates CONTINUOUS (Na Nongkhai et al., 2022), a domain ontology that defines common concepts of multiple programming languages (Python, Java, and C#), to support learners in understanding programming concepts. In addition, ADVENTURE employs the Elo rating system (ERS) in an educational setting (Pelánek, 2016) as an adaptive mechanism to recommend exercises appropriate for individual learners' skill levels.

In ADVENTURE, learners first select a programming language and one concept, and complete a pre-test to determine their initial difficulty level (Easy, Standard, or Difficult); then they will be provided with exercises in response to their skills. The system updates their skill based on their code submissions and recommends the next exercises. Once the threshold at the Difficult level, learners progress to the next concept suggested by CONTINUOUS or through their own selection. For the code assessment, ADVENTURE automatically assesses learners' code using unit tests by comparing code outcome with the test cases that are defined for each question. As a result, learners received only correctness information for each test case rather than explanatory feedback. Such a limitation constrains the system's capacity for personalized learning, as it does not clarify the reasons for incorrect solutions or guide learners on how to improve their code structure.

To address this limitation and further support learners in enhancing programming skills, this research integrates RAG and GenAI into ADVENTURE, to improve the functionality of our system, such as the assessment of learners' code, the provision of automatic feedback on learners' coding answers, and the recommendation of the next suitable exercise. Specifically, by incorporating the GPT-4 model, a well-known LLM at the time of the system's development, within a RAG-based feedback framework, ADVENTURE can retrieve relevant information from both the CONTINUOUS ontology and learning records before generating explanations and suggestions. This integration aims to support learners' comprehension of feedback on their coding answers and compare the learner preferences to next-exercise recommendations suggested by the adaptive approach and GenAI. Therefore, the following three research questions will be explored:

1. How can RAG be effectively integrated into an ontology-based adaptive learning system to support formative feedback in programming education?
2. What are the differences in learning performance, including achievement and perception, among learners using (1) an adaptive approach without GenAI, (2) an adaptive approach with GenAI feedback (using GPT-4) and exercise recommendations, and (3) a hybrid approach combining adaptive and GenAI-driven recommendations with GenAI feedback?
3. How do learners perceive the helpfulness of feedback provided by GenAI?

To address the research questions, we developed three distinct modes within ADVENTURE: an adaptive mode, a GenAI mode, and a mode integrating the adaptive and GenAI modes. An experiment involving 87 undergraduate students in a Thai university was conducted, and data on learning perception and learning logs were collected and analysed.

Our primary contribution is the integration of knowledge graphs and user interaction history with RAG, establishing a more capable and reliable foundation for LLMs in education. To the best of our knowledge, this represents one of the first studies to design and evaluate such an integrated framework specifically for intelligent educational systems. Complementing this architectural innovation, we provide novel, data-driven insights through a comparison of real-world learning performances in programming practice across varied recommendation strategies and feedback mechanisms. These empirical results not only validate our approach but also offer valuable guidance for the future development of personalized, LLM-based education.

The remainder of this paper is organized as follows: Section 2 reviews related work on the generation of automatic feedback on programming code using LLMs, and the integration of RAG and LLMs in educational systems. Section 3 presents the design and development of the three distinct modes within ADVENTURE. Section 4 describes the experimental settings employed in this study. Section 5 reports the results, including analyses of questionnaire responses and learning logs. Finally, Section 6 concludes the paper by discussing key findings, limitations, and directions for future work.

## 2. Literature review

### 2.1. LLMs in programming and educational contexts

The rapid growth of LLMs in GenAI has prompted numerous studies to explore their application in designing and developing instructional strategies within programming education. For instance, Azaiz et al. (2024) investigated the application of LLMs for generating automated feedback on students' programming exercises. Moreover, several studies have investigated the potential of AI chatbots, such as ChatGPT and Gemini, to create instructional strategies that effectively support learners in practicing programming (Kosar et al., 2024; George & Dewan, 2024; Majumdar et al., 2024).

Furthermore, Zhang and Liu (2024) introduced JavaLLM, a model fine-tuned from CodeQwen 1.5 using Java-specific resources such as documentation, assignments, community discussion, and code feedback. JavaLLM demonstrated improved performance in code generation and question answering compared to its base model, proving effective in classroom-oriented Java learning. This highlights the benefits of domain-specific LLMs for enhancing instructional quality and personalized learning. However, a limitation of this model was that it facilitated Java education only. Kazemitabaar et al. (2024) introduced CodeAid, an LLM-based programming assistant powered by GPT-3.5, designed to provide pedagogical support in programming education without offering direct code solutions. CodeAid was deployed in a 12-week university-level C programming course, and assisted with debugging, conceptual explanations, and pseudocode scaffolding. Learners engaged with CodeAid during both in-class activities and homework assignments, reporting high response accuracy (79 %) and perceived helpfulness (86 %). Educators highlighted CodeAid's scalability and alignment with academic integrity, while students appreciated its private, context-aware feedback. This study demonstrated the potential of controlled LLM integration for ethical, personalized programming support in formal education. Tang et al. (2024) introduced SPHERE, an interactive system that utilized GPT-4 to support instructors in delivering scalable, personalized feedback for in-class coding exercises and peer discussions. Implemented in a large Python course, SPHERE generated feedback drafts, which highlighted issues, strategies, and examples, for instructors to review and customize. This study showed that GPT-4 produced high-quality, context-sensitive feedback without increasing instructors' workload which emphasized SPHERE's potential to enhance personalized learning.

Beyond classroom-based systems, numerous studies have integrated LLMs in the development of personalized learning tools for online education. Yang et al. (2024) proposed CREF, a conversational repair framework designed to support programming tutors by leveraging LLMs to generate and iteratively refine code corrections. They developed TutorCode, a large-scale dataset comprising 1239 incorrect C++ code samples annotated with tutor guidance, solution descriptions, and failing test cases. Using this dataset, the authors conducted experiments with 12 LLMs and found that GPT-4 provided with instructional context, achieved the highest correctness rate (76.6 %). CREF's multi-stage approach, using dialog history and contextual prompts, enhanced repair quality while reducing tutor's debugging workload by 71.2 % and cutting response time from 26.7 to 7.7 min. These findings highlight the potential of conversational LLMs to deliver scalable, high-quality feedback in programming education. Gabbay and Cohen (2024) proposed a pedagogical approach in an online programming MOOC (Massive Open Online Course) by combining automated test-case results with GPT-generated explanations, aiming to improve learners' understanding and problem-solving skills. Conducted within a large introductory Python course, the approach was evaluated through analysis of learner interactions and performance. Their findings indicated that learners who engaged with the combined feedback showed improved correction rates and greater persistence in debugging their code. Additionally, learners reported high levels of perceived usefulness and clarity regarding the feedback. This study underscores the effectiveness of integrating LLM-generated feedback with automated assessment to support scalable, personalized learning in programming education. Similarly, the study by Li et al. (2024b) introduced ProgMate, an intelligent programming assistance that integrates LLMs with knowledge graph reasoning to support personalized learning in large-scale programming courses. By analyzing incorrect code through LLM-guided abstract syntax tree (AST) representations, the system identifies learner difficulties and maps them to relevant concepts within a knowledge graph. ProgMate employs probabilistic skill modeling to assess proficiency and generate adaptive feedback, guiding learners along personalized paths based on their understanding, and the conceptual structure of C programming. Their study reported high predictive accuracy and positive learner feedback, particularly when incorporating GPT-4. These results demonstrate the promise of combining LLMs with structured educational frameworks to enhance scalable programming instruction.

As discussed above, several studies have integrated LLMs to provide formative feedback with the aim of reducing instructors' workload, such as SPHERE (Tang et al., 2024) and CREF (Yang et al., 2024). These systems employed LLMs as assistants to generate feedback based on learners' code and offered suggestions for instructors to review before delivering them to students. Conversely, other studies, such as CodeAid (Kazemitabaar et al., 2024) and Gabbay and Cohen (2024), focused on supporting learners by leveraging GenAI to provide feedback that promotes learning while avoiding direct code solutions. Therefore, this research aims to enhance ADVENTURE by utilizing LLMs to automatically generate feedback and recommend the next programming exercise for learners. While ProgMate (Li et al., 2024b) offers individualized learning pathways and personalized feedback, ADVENTURE incorporates an ontology of multiple programming languages (referred to as CONTINUOUS) and employs an adaptive mechanism using the ERS in an educational setting. This mechanism works alongside LLM-generated feedback produced by the GenAI model (GPT-4).

*2.2. Retrieval-augmented generation techniques and knowledge graphs for LLMs*

While previous studies have primarily focused on how LLMs can support learning programming, few have examined the underlying technical mechanisms that enable the personalized feedback. As shown in section 2.1, LLMs have demonstrated significant potential in programming education due to their ability to understand semantic context and generate high-quality text responses (Gan et al., 2023), allowing them to act as intelligent virtual tutors, providing instant and personalized feedback. Moreover, recent studies show that LLMs exhibit strong interdisciplinary reasoning, which is valuable in addressing complex educational scenarios (Xu et al., 2024a). However, the application of LLMs in education presents significant challenges. LLMs are prone to hallucinations where answers are produced that appear plausible but are factually incorrect, which may mislead learners (Li et al., 2024a). Enhancing factual reliability and adaptability remains a crucial direction for educational AI research.

To mitigate these limitations, retrieval-augmented generation (RAG) has emerged where an LLM's internal knowledge is combined with an external data source, allowing it to retrieve relevant, up-to-date information during response generation (Lewis et al., 2020). This architecture typically includes a retriever, which locates the most pertinent documents, and a generator, which synthesizes them with the query to produce context-aware responses. By grounding generation in retrieved evidence, RAG reduces hallucinations and improves factual accuracy (Lewis et al., 2020). In education, Dong et al. (2023) developed an AI tutor using LangChain (Chase, 2022) that organizes course materials into a knowledge base and uses RAG to retrieve relevant information for student queries. The retrieved data is combined with the question to form a new prompt, which the LLM processes to generate responses.

Beyond retrieval-based approaches, Knowledge Graphs (KGs) have been increasingly adopted to structure educational knowledge and enhance personalized learning. Earlier work used a course-centered ontology to link knowledge points to existing language learning materials (Wang et al., 2014), whilst more recent studies introduce ontology-driven reinforcement learning frameworks that adapt teaching strategies based on student performance and provides personalized learning suggestions (Hare & Tang, 2024). By explicitly modelling conceptual structures, KGs enable more interpretable and targeted feedback for learners.

Integrating KGs with RAG further enhances reasoning and retrieval capabilities. Traditional RAG struggles with complex, multi-hop reasoning and unstructured data, leading to degraded retrieval accuracy in long or conceptually dense queries (Chang & Zhang, 2024; Xu et al., 2024b). KG-enhanced RAG addresses these challenges by providing structured, relational context to guide the retrieval process and strengthen logical consistency. Recent approaches incorporate community-based graph structures or parsed historical data to improve retrieval precision and fact-checking (Chang & Zhang, 2024; Xu et al., 2024b). This integration is particularly relevant in educational contexts, where learning materials often form interconnected knowledge networks.

In summary, the convergence of LLMs, RAG, and KGs presents a powerful foundation for building intelligent educational systems. RAG enhances factual accuracy, KGs provide structure and interpretability, and their integration enables adaptive, context-aware learning support. Building on these advances, ADVENTURE integrates a KG within a RAG process to find and recommend exercises that are closely linked to what the learner is currently having difficulties with. The system uses the learner's session history to help make recommendations during their practice, and detailed log data for later analysis of learning behavior is also collected. Furthermore, we explore the differences in learning performance across three distinct strategies, two of which are then integrated with GPT-4. This combination of KG-driven RAG, session-aware recommendations, and adaptive mechanisms distinguishes ADVENTURE's design, offering a novel approach to personalized programming practice. The details of each strategy implemented within ADVENTURE is described in Section 3.

## 3. The design and development of the three modes in ADVENTURE

This study builds upon the previously developed ADVENTURE system (Na Nongkhai et al., 2023, 2025), which provides programming exercises based on learners' skills. To address the research questions, ADVENTURE was extended by implementing three distinct modes:

(1) An adaptive mode, named ADVENTURE-2.0, which evaluates learners' code using unit tests that compare the program's output with the expected results and provide feedback (by unit test) based on the outcome. Moreover, learners' skills are updated using the Elo rating system (ERS) (Pelánek, 2016), allowing the system to recommend exercises based on each learner's current skill.
(2) Two GenAI-based modes with different recommendation strategies which both employ a GPT-4 model and CONTINUOUS as a KG to evaluate learners' code and generate formative feedback based on their submissions. (a) *A GenAI mode*, named ADVENTURE-3.0, uses GPT-4 to recommend the next appropriate programming exercise according to the learner's performance. Only if the recommended exercise has previously been successfully completed, an alternative exercise recommended by the adaptive mode is provided. (b) *A hybrid GenAI-adaptive mode*, named ADVENTURE-3.1, provides learners with the option to choose between the GenAI or the adaptive mode recommendations. In this mode, learners can reject the GPT-4 recommendation and instead choose the exercise recommended by the adaptive mode. However, if they choose the recommendation of the GenAI model and the recommended exercise happens to be a repeated one, the learner is reminded and provided an option to change their decision to use the recommendation of adaptive mode.

The GenAI process of the latter two modes is initiated after a learner submits their code, as shown in Fig. 1. The system sends submission information, including a personalized identifier, the programming question, programming language, the learner's code, and the users chat history from the front end to the back end as part of the prompt data for the GenAI model. Two prompts were designed to meet the specific requirements of the RAG process, as shown in Figure A1. The first prompt is the composite prompt, which verifies the correctness of the submitted code, identifies errors, provides corrected solutions, and explains related misconceptions. It also recommends new exercises based on the learner's past performance to address specific knowledge gaps identified in the users' skills. The second prompt, the question reformulation prompt, transports potentially ambiguous queries into clear, standalone questions, ensuring that retrieval and generative responses are contextually accurate.

The prompt data also includes a KG, which serves as the foundation for the system's RAG process. This graph, built from the CONTINUOUS ontology of multiple programming languages, represents the programming question as a node and captures the relationships between questions. By leveraging this graph, the system identifies relevant exercise nodes closely related to the learner's current problem. These nodes are stored in a vector database, which supports efficient semantic search and retrieval of exercises most appropriate to the learner's current progress. The graph can be updated using the Cypher graph query language to query, modify and expand nodes within the database.

The system constructed using the LangChain framework (Chase, 2022), then decomposes the learner's query into multiple components, integrating both the learner's input and their interaction history via a history-aware retriever. This retriever accesses the vector store to select exercises and information that are contextually relevant based on the learner's past submissions and performance. After verifying the correctness of the submitted code, the system provides the retrieved context to a LangChain Question-Answer Chain, which then generates formative feedback, including explanations of errors when present. Using the KG and the learner's history, the system then recommends subsequent exercises that target specific knowledge gaps while avoiding unnecessary repetition.

The entire interaction history is stored in a memory component, allowing the system to maintain context over multiple sessions and provide progressively personalized support. This approach, combining KG-driven retrieval with memory-enhanced feedback, offers a novel and effective way to deliver adaptive, context-aware programming practice tailored to each learner's evolving needs.

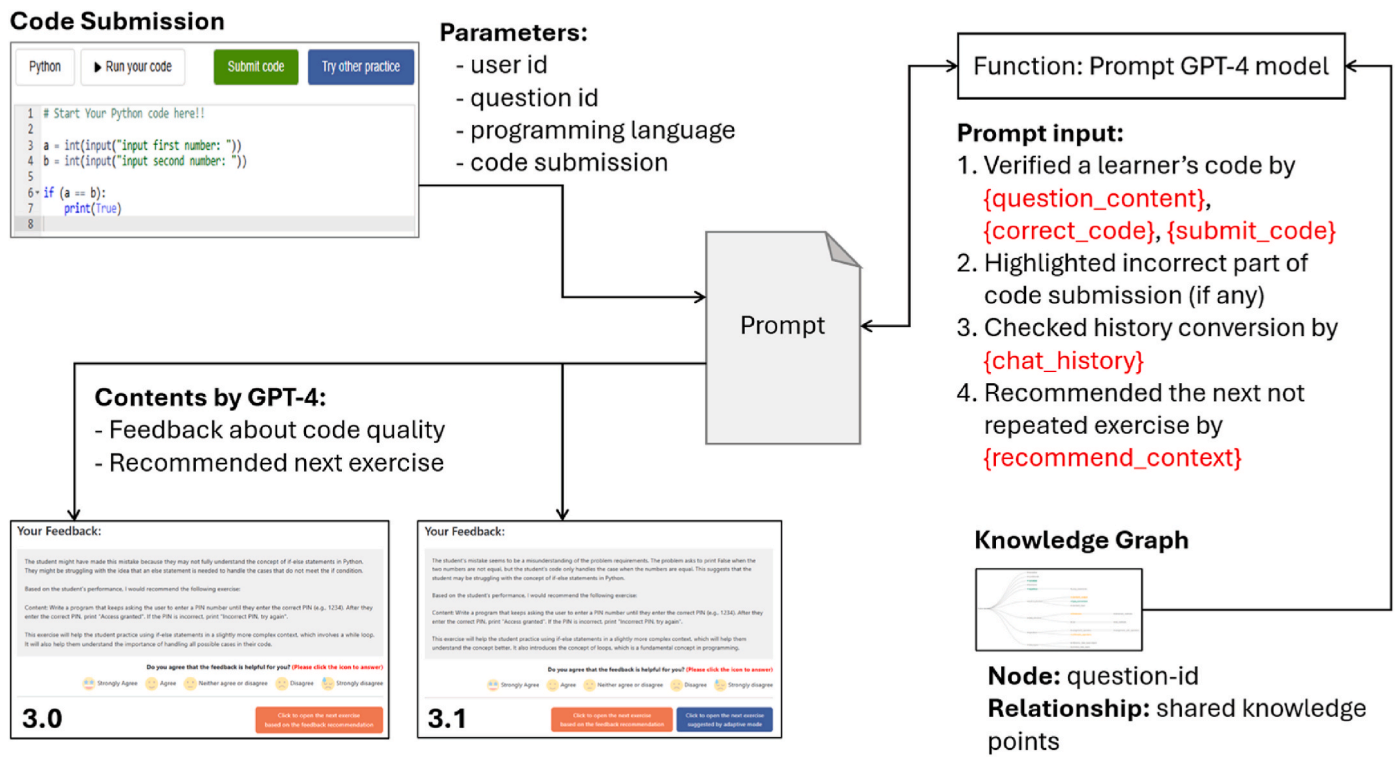

**Fig. 1.** Integration of GenAI within ADVENTURE.

## 3.1. An adaptive mode (ADVENTURE-2.0)

The adaptive mode (ADVENTURE-2.0) integrates several enhancements based on ADVENTURE-1.1 (Na Nongkhai et al., 2025). For instance, the updated system features: (1) a progress bar to track learners' skills at each difficulty level (as shown in a highlight box in Fig. 2); (2) expanded hints with additional knowledge points for each question (as displayed under "Hints for this question:" in Fig. 2 and Fig. 3); and (3) the display of only the incorrect test cases to learners.

In the initial process of ADVENTURE for all modes, after logging in, learners select a programming language and concept to practice. If the selected concept has not been previously attempted, the system prompts them to complete a pre-test to determine their initial difficulty level. Based on this result, learners are directed to begin exercises at the assigned level. These steps serve as the entry point for all modes implemented in ADVENTURE.

For the specific processes in this mode, which consist of four steps used to identify the next suitable exercise. These steps are initiated after learners submit their code to the system as follows: (1) evaluate the submitted code using unit tests; (2) update learners' skills and the difficulty of the attempted exercise; (3) identify an exercise with a difficulty level closely matching the learner's current skills; and (4) recommend the selected exercise to the learner. Through this cycle, both learners' skills and exercise difficulty levels are continuously updated with each submission, ensuring that the next exercises remain aligned with the learner's current skills.

To assess a learner's code in this mode, we employ a unit testing approach that verifies test cases in each programming question. Fig. 3 shows an example of a programming task. If the learner submits incorrect code, for example by writing the wrong condition and missing an else case (as shown in Fig. 4), and therefore fails to resolve any test cases for the current question, the system displays those failed test cases to the learner, as shown in Fig. 5.

In contrast, if the code successfully passes all test cases, the system provides a message instructing the learner to proceed to the next exercise, as shown in Fig. 6. In this mode, learners receive exercises recommended by the adaptive mechanism in the system. For feedback on code submissions, the system displays only the failed test cases and provides limited explanations. To address this limitation, formative feedback generated by GenAI is designed and implemented.

## 3.2. A GenAI mode (ADVENTURE-3.0)

The GenAI mode (ADVENTURE-3.0) was developed based on ADVENTURE-2.0 by integrating the GPT-4 model (Fig. 1) into the system. In this mode, an API was developed to connect the front-end developed with React (https://reactjs.org/, accessed on October 1, 2022), with the back-end built with Python, to call the GPT-4 model (OpenAI, 2023). After receiving a response from the GPT-4 model, the front-end displays two pieces of content ((1) feedback on code quality and (2) a recommended exercise) on the feedback interface, as shown on the left on Fig. 7.

In the feedback interface for code submissions, the code feedback generated by the GenAI model is highlighted in the first box in Fig. 7, while the content of recommended next exercise highlighted in the second box in Fig. 7. Before proceeding, the system also requires the learner to confirm their agreement with this feedback. Compared to the feedback presented in ADVENTURE-2.0 (Fig. 5), the feedback in ADVENTURE-3.0 offers more detailed explanations to support learners in identifying and understanding concepts they find difficult. Additionally, it provides specific details regarding the recommendations for the next exercise.

Nevertheless, there are some limitations to the exercise recommendations provided by the GenAI model in this mode. Even though the chat history was prompted to avoid recommending repeated questions, GPT-4 still sometimes suggests a question that learner has already completed. In this case, a pop-up dialog prompts learners to indicate whether they wish to repeat the recommended exercise. If learners choose to repeat it, the system reassigns this exercise to learners; otherwise, the system provides a new exercise selected by the adaptive mechanism. The adaptive process serves as an alternative path within the GenAI mode, invoked only in this specific scenario when a repeated exercise is recommended.

## 3.3. The integration of adaptive and GenAI modes (ADVENTURE-3.1)

The hybrid GenAI-Adaptive mode (ADVENTURE-3.1) was developed based on ADVENTURE-3.0 by combining the adaptive recommendation and GenAI recommendation. As shown in Fig. 7, the main difference from 3.0 is that in 3.1, the adaptive model is always available as an option, not only when GPT-4 recommends a previously completed exercise. In this mode, learners have the options to select the next programming exercise recommended either by the GenAI or the adaptive approach (by clicking the blue button).

In other words, ADVENTURE-3.1 introduces a recommendation process where learners first see the GenAI model's recommended exercise. If they do not agree with the GenAI recommendation, then they are able to use the adaptive mode recommendation as in ADVENTURE 2.0. Learners only see the adaptive recommendation after declining the GenAI recommendation. If the learner also rejects the adaptive mode's recommendation, the system will provide an alternative exercise recommended by the adaptive mode.

If they accept the GenAI recommendation initially, the exercise is assigned as in ADVENTURE 3.0. In this case, only if they have already completed the exercise, the learner will be reminded if they want to get another recommendation by adaptive mode, otherwise they will be provided with a GenAI recommendation. ADVENTURE-3.1 integrates both the adaptive and GenAI suggestions into the recommendation strategy and offers the learners a more flexible approach, where they can choose between either recommendation.

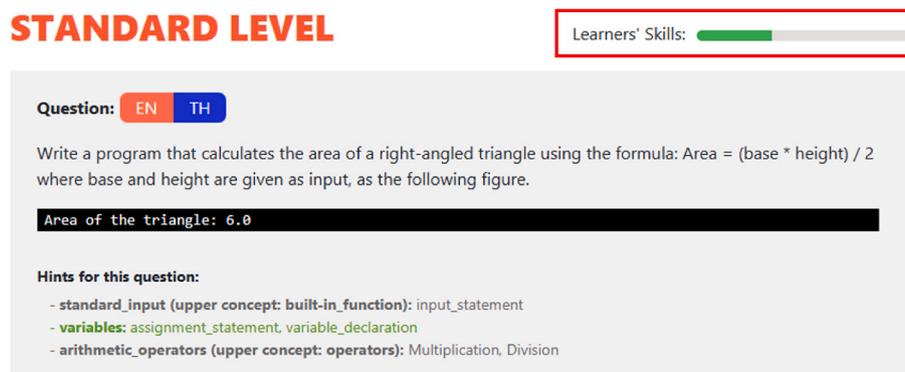

**Fig. 2.** Display of a learner's skills progress bar in each difficulty level (as highlighted in the red box).

Fig. 3. An example of a programming question.

Fig. 4. An example of a learner's code.

Fig. 5. An example of failed test cases.

Fig. 6. The message after passing all test cases.

## 4. Experimental description

### 4.1. Participants and measurement techniques

87 undergraduate students from the first and second years, majoring in Information Technology at a university in Thailand participated in the experiment. The first-year students had prior experience with one programming course, consisting of 30 h of lectures and 45 h of practical activities. The second-year students had completed two programming courses, each consisting of 30 h of lectures and 45 h of practical activities. To guarantee fairness, participants were assigned to three experimental groups where each group consisted of 15 first-year and 14 second-year undergraduate students, totaling 29 participants per group. The groups were organized as follows: (1) Group A used the adaptive mode (ADVENTURE-2.0), consisting of 23 males and 6 females; (2) Group B used the GenAI mode (ADVENTURE-3.0), consisting of 22 males and 7 females; and (3) Group C used the GenAI-adaptive mode (ADVENTURE-3.1), consisting of 28 males and 1 female. Since this study targeted novice learners with limited programming experience, achieving balanced gender representation across groups was challenging. This limitation is discussed in Section 6.

Fig. 7. Examples of code submissions' feedback and the next exercise recommendation in ADVENTURE 3.0 and 3.1.

This study employed two measurement techniques: (1) learning logs including the learning outcome of each programming exercise, learner perception of each piece of feedback provided after code submission (only for Groups B and C), and other learning behaviors, and (2) a learning perception questionnaire. This questionnaire was developed in Thai based on (Wang et al., 2014, 2020, 2024), with additional guidance from experienced instructors to evaluate learners' perceptions of their learning experience. It comprised 26 items measured on a seven-point Likert scale, where scores of 1–3 indicated varying degrees of disagreement, 4 represented a neutral response, and 5 to 7 indicated varying degrees of agreement. The items were categorized into four dimensions: (1) mental effort, reflecting effort to understand the purpose of the experiment and perform the learning activities; (2) mental load, reflecting distraction and pressure; (3) technology acceptance, encompassing ease of use and perceived usefulness; and (4) satisfaction. All participants completed the questionnaire after the learning activities.

### 4.2. Experimental procedure

Fig. 8 presents the experimental procedure employed in this study. All experimental groups were given 15 days to use the ADVENTURE system in their assigned mode. Upon selecting a programming concept, the system provided participants with a pretest consisting of three questions, each representing a difficulty level). If a participant failed a question, they were assigned to the corresponding difficulty level of that question for subsequent programming exercises.

In addition to allowing access to the system at the convenience of the participants, on-site sessions were conducted on three separate days (on the 1st, 7th, and 15th day of the experimentation period). Each session lasted 1.5h and was intended to monitor participants' learning progress while using the system in a controllable environment. On the 1st day, each participant was provided with a pre-assigned system account and was guided through the initial login process and system usage. On the 7th day, participants' activity was monitored again. This time they were encouraged to ask for support when there were bugs or technical difficulties while using the system. On the 15th day, participants were asked to complete a questionnaire to provide feedback on their learning experience. All other sessions throughout the experimentation period were not monitored in a classroom and the participants were encouraged to use our web-based system during their free time outside the classroom.

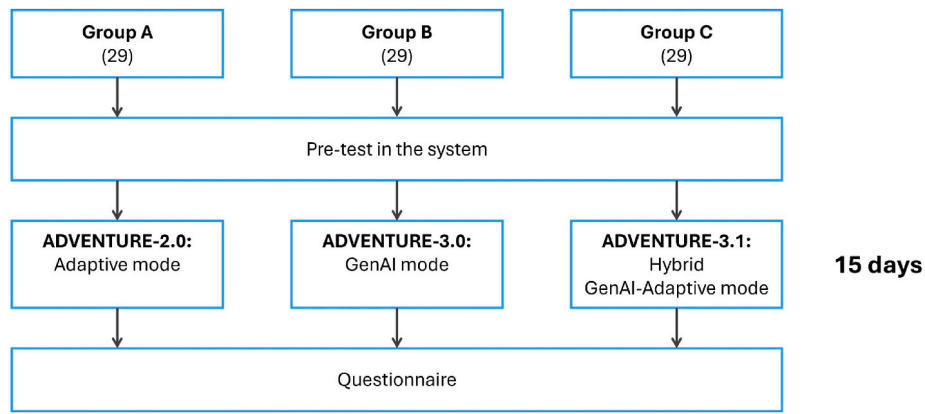

Fig. 8. Experimental procedure.

## 5. Results

### 5.1. The analysis of questionnaire responses

Table 1 presents a summary of the analysis results of the responses from 87 participants across all groups to the learning perceptions questionnaire. Each participant was required to complete 26 questions, which were organized into four categories: mental effort (2 questions, $\alpha = 0.582$); mental load (4 questions, including 3 questions related to distractions from three system features and 1 question measuring pressure, $\alpha = 0.685$); technology acceptance (10 questions covering ease of use and usefulness of five system features $\alpha = 0.853$); and satisfaction (10 questions, $\alpha = 0.904$). Additionally, Harman's single-factor test indicated the first factor accounted for 37.70% of the total variance, which is below 50%. Therefore, common method bias was not considered a serious issue in this study.

For "mental effort," the average rating for understanding the purpose of using the system across all experimental groups approached the neutral point, at 4.62. Standard deviations were slightly higher in the group who used the GenAI mode. This suggests that all participants perceived a comparable, neutral level of mental effort in understanding the purpose of using ADVENTURE, and that this perception was consistent across the different modes. Meanwhile, the average rating for the learning activity item across all groups exceeded the neutral point, with means of 5.48 (S.D. = 1.06) for Group A, 6.24 (S.D. = 1.27) for Group B, and 5.52 (S.D. = 1.21) for Group C. These results suggest that participants in the GenAI mode (Group B) exerted greater effort in practicing programming activities compared to those in the other modes. Furthermore, ANOVA results suggest that there is a significant difference among the groups (F (2,84) = 3.795, $p < 0.05$, $\eta^2 = 0.083$). Post hoc pairwise comparisons indicated that Group B required significantly higher mental effort in the learning activity than Group A ($\Delta M = 0.759$, $p = 0.017$), and Group C ($\Delta M = 0.724$, $p = 0.022$).

Regarding "mental load," the average distraction ratings for Group A and Group B were slightly below the neutral point, at 3.86 (S.D. = 1.52) and 3.75 (S.D. = 1.47), respectively. In contrast, Group C reported a rating of 4.46 (S.D. = 1.39), slightly above the neutral point, suggesting participants in Group C perceived a slightly higher distraction while using the system. ADVENTURE-3.1 required learners to select between GenAI and adaptive modes for the next exercise, which may have added complexity and contributed to the slightly higher distraction ratings in Group C. Meanwhile, the average pressure ratings for Group B and Group C were 5.55 (S.D. = 1.57) and 5.17 (S.D. = 1.61), respectively, while Group A reported slightly lower pressure (Mean = 4.90, S.D. = 1.84). This indicates that participants who used the system integrated with GenAI may have experienced some challenges when practicing programming, possibly through exercises recommended by the GenAI model, which leads to higher perceived pressure. However, ANOVA results suggest that there are no significant differences among all groups.

Regarding "technology acceptance," the average ratings for ease of use across the five system features, including the graph visualization of programming concepts, the programming exercise webpage, the hints in each programming question, the feedback addressing code submissions, and the next programming exercises suggestion, are more than 5 for all three groups. (We will discuss the rating of the last two features in more detail in Section 5.3.) Similarly, the average rating for usefulness also exceeded 5 for all groups. These results suggest that the system is easy to use and useful for practicing programming, regardless of the mode in which it is used. Moreover, ANOVA results suggest there are no significant differences between all groups. Consistent with this finding, the average rating for general satisfaction exceeded 5 across all groups. Furthermore, ANOVA results suggest that there are no significant differences between all groups. These results suggest that participants across all groups found ADVENTURE generally satisfying for programming practice.

### 5.2. The analysis of learning logs

During the experimentation period, learners' activities in ADVENTURE were automatically recorded as learning logs. These logs comprised responses to the programming experience questionnaire completed at first login, pre-test results for each programming concept, learners' skills, and code submissions. A post-hoc analysis of baseline programming proficiency between first- and second-year students showed only a minimal difference in prior programming knowledge, with the first-year students achieving an average pre-test score of 51.35% and the second-year students achieving 53.53%. Among these, a total of 4956 code submissions were collected from all participants: 1633 from Group A, 2030 from Group B, and 1293 from Group C. For the analysis of learning outcomes, the code submissions were categorized into four distinct features: (1) the submission of correct answers; (2) the submission of wrong answers; (3) the submissions lacking essential programming logic; and (4) the frequency of requesting another exercise. Fig. 9 shows a summary of these four features in learning logs across the three experimental groups.

For the "submission of correct answers," as shown in (1) of Fig. 9, the bar chart indicates that Group C achieved the highest average number of correct submissions. Furthermore, the ANOVA results suggest that there were significant differences among all groups (F (2,84) = 6.343, $p < 0.01$, $\eta^2 = 0.131$). Post-hoc comparisons showed that Group A had significantly fewer correct submissions than Group B ($\Delta M = 0.091$, $p < $

**Table 1**
Summary of questionnaire responses on learning perceptions.

| | | Mental Effort | | Mental Load | | Technology Acceptance | | Satisfaction |
|---|---|---|---|---|---|---|---|---|
| | | Purpose | Learning Activity | Distraction | Pressure | Ease of use | Usefulness | |
| Group A (ADVENTURE 2.0) | Mean | 4.62 | 5.48 | 3.86 | 4.90 | 5.59 | 5.72 | 5.34 |
| | S.D. | 1.63 | 1.06 | 1.52 | 1.84 | 0.72 | 0.72 | 0.75 |
| Group B (ADVENTURE 3.0) | Mean | 4.62 | 6.24 | 3.75 | 5.55 | 5.57 | 5.66 | 5.37 |
| | S.D. | 1.90 | 1.27 | 1.47 | 1.57 | 0.85 | 0.89 | 1.02 |
| Group C (ADVENTURE 3.1) | Mean | 4.62 | 5.52 | 4.46 | 5.17 | 5.63 | 5.77 | 5.38 |
| | S.D. | 1.63 | 1.21 | 1.39 | 1.61 | 0.97 | 0.82 | 1.02 |
| One-way ANOVA | F | 0 | 3.795 | 1.994 | 1.117 | 0.042 | 0.136 | 0.016 |
| | P | 1 | 0.026* | 0.142 | 0.332 | 0.959 | 0.873 | 0.984 |

Note: df(2, 84) in ANOVA, **$p < 0.01$, *$p < 0.05$.

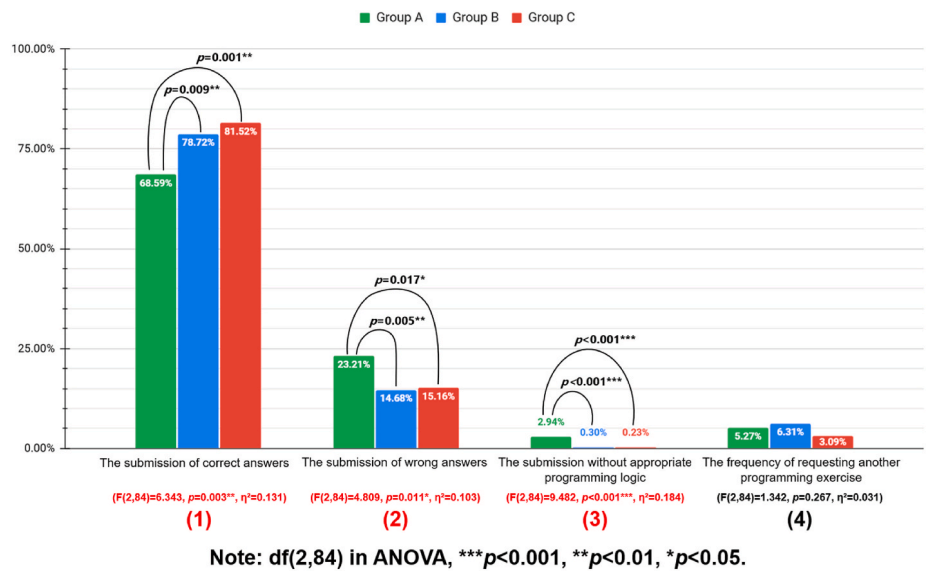

**Fig. 9.** Comparison of four learning log features among the three modes within ADVENTURE: (1) The submission of correct answers; (2) The submission of wrong answers; (3) The submission missing appropriate programming logic; and (4) The frequency of requesting another programming exercise.

0.01), and Group C ($\Delta M = 0.114$, $p < 0.01$). However, there were no significant differences between the GenAI and hybrid modes suggesting that both modes outperformed the adaptive mode in correct code submission rates, with the hybrid mode showing a slightly higher mean difference than the GenAI mode, indicating that the hybrid mode may provide more substantial benefits for programming practice than the other modes.

For the "submission of wrong answers," as shown in (2) of Fig. 9, the bar chart indicates that Group A submitted the highest number of incorrect answers compared to the other groups. The ANOVA results suggest that there were significant differences across all groups (F (2,84) = 4.809, $p < 0.05$, $\eta^2 = 0.103$), and a pairwise comparison showed that Group A had significantly more incorrect submissions than Group B ($\Delta M = 0.070$, $p < 0.01$), and Group C ($\Delta M = 0.059$, $p < 0.05$). This finding suggests that GenAI-based feedback and exercise recommendations may have helped participants more effectively identify and correct their mistakes compared to the adaptive mode alone. However, there were no significant differences found between Groups B and C, possibly due to both groups using a system that incorporated GenAI.

For the "submission of code missing appropriate programming logic," which refers to scenarios in which participants submitted either empty code or code that lacked the essential logic required to align with the task objectives, regardless of whether the output appeared correct. This type of submission differs from incorrect code submissions, which typically reflects an attempt to solve the problem but results in an incorrect outcome. As illustrated in (3) of Fig. 9, the bar chart indicates that Group A submitted a greater number of code submissions lacking the essential programming logic compared to the other groups. The ANOVA results suggest that there were significant differences across all groups (F (2,84) = 9.482, $p < 0.001$, $\eta^2 = 0.184$). These findings suggest that participants who engaged with the system using the adaptive mode were more likely to submit code missing the essential programming logic than those who used the other two versions incorporating the GenAI model. This is supported by pairwise comparisons between Group A and Group B ($\Delta M = 0.040$, $p < 0.001$) and between Group A and Group C ($\Delta M = 0.041$, $p < 0.001$), both of which showed significant differences.

For the "frequency of requesting another programming exercise," which refers to instances in which participants abandoned the current exercise and opted to try a different one instead, the ANOVA result suggests that there were no significant differences among the groups. However, as shown in (4) of Fig. 9, Group C in average requested alternative exercises less frequently than participants in the other groups. This suggests that participants using the integration of adaptive and GenAI modes may have received more appropriate exercise recommendations than those using the GenAI or adaptive mode alone, which was possible because participants had the option to select the next exercise recommended by either the GenAI or the adaptive mode.

In summary, the overall results (Fig. 9) indicate that participants in Group C submitted a higher number of correct answers and showed fewer instances of submissions missing essential programming logic, as well as fewer requests for alternative exercises, compared to the other groups. This suggests that providing more comprehensive feedback in response to code submissions, as well as offering the option to choose the next programming exercise recommended by GenAI or the Adaptive algorithm, may enhance the efficiency of programming practice activities.

*5.3. The analysis of learners' responses with the feedback generated by GenAI*

To address the third research question in this study, we collected and analysed learners' responses to the question: "Do you agree that the feedback is helpful for you?", as illustrated in the bottom of Fig. 7. A total of 3154 responses were collected from participants in Group B and Group C.

The bar chart in Fig. 10 presents the level of agreement with feedback generated by the GenAI model, comparing two experimental groups using a five-point Likert scale ranging from Strongly Agree (5) to Strongly Disagree (1). The average rating of Groups B and C was 4.01 (S. D. = 1.27), suggesting that most participants agree that the feedback created by GPT-4 was helpful. Moreover, a *t*-test indicated that there were no significant differences between Group B (Mean = 4.27, S.D. = 0.91) and Group C (Mean = 4.15, S.D. = 0.87, t (56) = 0.831, p = 0.409).

Table 2 presents learners' perceptions within the technology acceptance dimension of two system features: feedback on code submissions, and recommendation for the next programming exercise. For the feature of "feedback on code submissions," the average ratings across all groups were above 5 for both ease of use and usefulness, indicating that participants slightly agreed that this feature was easy to use and beneficial. Furthermore, the qualitative feedback, as shown in Table 3, revealed that while the simple feedback generated by unit tests in ADVENTURE-2.0 was perceived as easy to understand and helpful (AP-1, AP-2), some participants required more detailed explanations of their coding mistakes and greater accuracy in the evaluation of their answers (AN-1, AN-2). In contrast, several participants in Group B and Group C provided positive comments that the feedback generated by GenAI was helpful (BP-1 to 3, CP-1 to 5). These findings suggest that although there was no significant difference in the average ratings across all groups, the qualitative feedback highlights the explanations generated by GenAI may support learner more effectively in understanding their code.

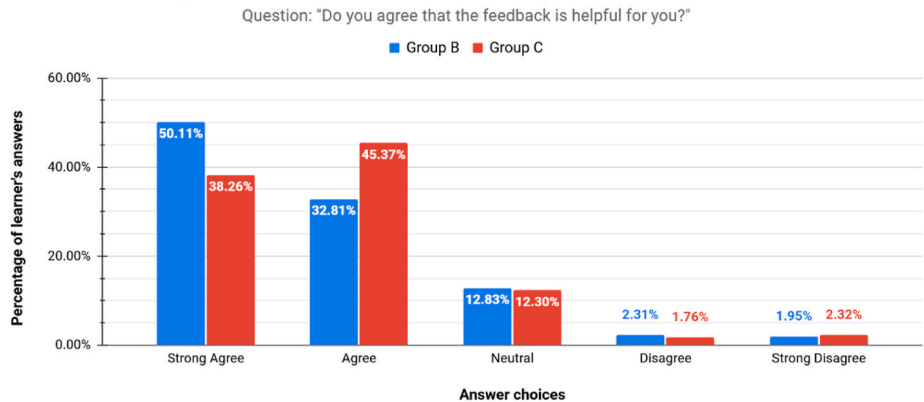

**Fig. 10.** Learners' responses to feedback generated by GenAI in ADVENTURE-3.0 and ADVENTURE-3.1.

**Table 2**
Technology acceptance for the features "Feedback on code submissions" and "Recommendation for the next programming exercise".

|  |  | Feedback on code submissions | | Recommendation for the next programming exercise | |
|---|---|---|---|---|---|
|  |  | Ease of use | Usefulness | Ease of use | Usefulness |
| Group A (ADVENTURE 2.0) | Mean | 5.38 | 5.76 | 5.48 | 5.45 |
|  | S.D. | 1.18 | 0.95 | 1.24 | 1.12 |
| Group B (ADVENTURE 3.0) | Mean | 5.14 | 5.69 | 5.59 | 5.48 |
|  | S.D. | 1.25 | 1.26 | 1.35 | 1.35 |
| Group C (ADVENTURE 3.1) | Mean | 5.38 | 5.76 | 5.41 | 5.55 |
|  | S.D. | 1.59 | 1.24 | 1.50 | 1.43 |

Regarding both ease of use and usefulness of the "Recommendation for the next programming exercise" feature, the average ratings of all three groups were greater than 5, indicating that participants slightly agreed that this feature was easy to use and beneficial. However, in the learning log data, Group B experienced a high rate of repeated exercise recommendations (76.95%) even GPT-4 was prompted not to recommend a repeated exercise. Furthermore, nearly half of the learners (31.69%) refused to repeat the same exercises that they already completed successfully. In this study, a repeated question is defined as an exercise that had previously been solved correctly by the learner but was subsequently recommended again by the system. In contrast, in Group C 71.69% of learners chose not to follow the GenAI recommendation and directly went for the adaptive suggestions. Similar to Group B, Group C also experienced a high rate of repeated exercise recommendations (71.54%) by GenAI. These quantitative findings are consistent with the qualitative feedback where participants in Group B (BN-1, BN-2) and Group C (CN-1, CN-2) noted frequent repetition of exercises.

Moreover, learners' qualitative comments were analysed and categorized into three types of feedback (Error Explanation, Code Suggestion, and Concept Clarification), as shown in Table 4. Most comments were identified as Error Explanation, indicating that learners valued feedback that identified and explained their mistakes, helping them recognize where their code went wrong. A smaller number of comments were classified as Code Suggestion, which provided hints or guidance toward improving code logic or achieving correct solutions. Additionally, some comments were categorized as Concept Clarification, suggesting that learners appreciated feedback that helped their understanding of programming logic beyond error correction. This highlights that the feedback may support learners in understanding their coding mistakes, correct solutions, and logic.

## 6. Conclusion and discussion

### 6.1. Discussion and implications

Recent research has increasingly explored the use of LLMs for generating automated programming feedback. For instance, Tang et al. (2024) introduced SPHERE, which integrated GPT-4 to generate feedback drafts to instructors before providing them to learners. Similarly, Gabbay and Cohen (2024) implemented GPT-generated explanation feedback in a Python MOOC, while Li et al. (2024b) proposed ProgMate, which used GPT-4 to provide formative feedback in C programming. These studies highlighted the effectiveness of a GPT-based model in delivering accurate and supportive feedback to enhance learners' programming practice.

Building upon these advantages, this study integrated GPT-4 into ADVENTURE within a RAG framework to address the first research question. The RAG design enabled GPT-4 to retrieve relevant conceptual and exercise information from ADVENTURE's ontology and learners' historical data before generating explanations, suggestions, and next-exercise recommendations. This approach reduced the likelihood of generic or context-insensitive responses, aligning feedback more closely with each learner's progress and prior activities.

To evaluate the effectiveness of this integration and answer the second research question, the study examined learning perceptions and learning logs across three approaches: (1) an adaptive approach without GenAI, (2) an adaptive approach with GenAI feedback (using GPT-4) and exercise recommendations, and (3) a hybrid approach combining adaptive and GenAI-driven recommendations with GenAI feedback.

In terms of learning perceptions, the results indicate that participants who engaged in the GenAI mode (Group B) experienced significantly more effort in practicing programming than those in the other groups. Additionally, Group B participants also reported experiencing more pressure during practice exercises suggesting that the exercises recommended by the GenAI model may have been more challenging for learners. In contrast, participants who engaged in the adaptive mode (Group A) reported the lowest level of effort and pressure, indicating that adaptive mechanisms may assign fewer challenging exercises to learners.

The analysis of the learning logs (Fig. 9) showed significant

**Table 3**
Learners' comments on ADVENTURE.

|  | Positive Comments | Negative Comments |
| --- | --- | --- |
| Group A (ADVENTURE 2.0) | AP-1: "The output is easy to understand." | AN-1: "It provides only output; I would like more explanation." |
|  | AP-2: "The function that highlights where the testcase is wrong is helpful." | AN-2: "Answer checking needs improvement." |
|  | AP-3: "The variety of exercises is great." |  |
| Group B (ADVENTURE 3.0) | BP-1: "The feedback is helpful." | BN-1: "The system loops through the same problems too often." |
|  | BP-2: "I like the guidance feature in the feedback." | BN-2: "I want the system reduce the frequency of repeated questions." |
|  | BP-3: "The post-code feedback provides excellent guidance and knowledge." |  |
|  | BP-4: "In terms of the suggestions after finishing the code, the advice and information given are really great." |  |
|  | BP-5: "In terms of the suggestions, that's the part I like the most, because it really helps me understand the logic better." |  |
|  | BP-6: "The system tells us where we went wrong after submitting the answer." |  |
| Group C (ADVENTURE 3.1) | CP-1: "The feedback after completing each exercise is very helpful." | CN-1: "The questions tend to repeat too often." |
|  | CP-2: "I like how it tells me exactly where I need to make corrections." | CN-2: "The orange button often repeats old questions." |
|  | CP-3: "I think it looks really good. It clearly shows where the program thinks something should be fixed or might be wrong." |  |
|  | CP-4: "Overall, I like that there's feedback to help understand where the mistake is." |  |
|  | CP-5: "I like how it tells me which parts I should fix." |  |

Note: Abbreviations are used to indicate the type of feedback and the participant group. P = Positive feedback, and N = Negative feedback. A = Students in Group A, B = Students in Group B, and C = Students in Group C. For example, AP refers to positive feedback from students in the group A, and AN refers to negative feedback from students in Group B.

**Table 4**
Type of feedback from learners' comments on ADVENTURE.

| Type of Feedback | Example Comment | Reasoning |
| --- | --- | --- |
| Error Explanation | BP-6, CP-2 to CP-5 | Identifying and explaining the learners' coding mistakes. |
| Code Suggestion | BP-3, BP-4, CP-5 | Providing suggestions or guidance toward the correct logic or an improved solution. |
| Concept Clarification | BP-5 | Supporting conceptual understanding rather than only correcting errors. |

Note: The example comment is taken from the learners' comments in Table 3.

differences among the groups across three features: correct submissions, incorrect submissions, and submissions without essential logic. Group A had significantly more incorrect submissions and submissions lacking essential logic than Groups B and C, and significantly fewer correct submissions. These findings suggest that participants using the GenAI modes practiced programming more effectively than those in the adaptive mode alone. Although there were no significant differences between Groups B and C, the bar chart showed that Group C achieved more correct submissions, fewer submissions lacking logic, and fewer skipped questions compared to Group B. This indicates that the hybrid GenAI–adaptive mode may provide more effective support for enhancing learners' programming skills than either mode alone.

Regarding learners' perceptions of feedback generated by the GenAI model, Yang et al. (2024) introduced CREF, which integrated GPT-4 to generate code feedback, and demonstrated that GPT-4 achieved the highest correctness rate among 12 evaluated LLMs. In this study, to examine the perceived helpfulness of feedback, we asked learners who engaged in both GenAI modes: "Do you agree that the feedback is helpful for you?" The analysis of learner's responses addressed the third research question: "How do learners perceive the helpfulness of feedback provided by GenAI?"

The results in Fig. 10 indicate that most participants in both Group B and Group C slightly agreed that the GenAI feedback was helpful in understanding their code submissions. Moreover, the results in Table 2 suggested that most participants slightly agreed that the two system features, code submission feedback and next exercise recommendations, were easy to use and useful. However, the learning log results revealed that many of the questions recommended in the GenAI mode had already been completed by participants. Similarly, the hybrid mode also frequently suggested repeated questions. The learner logs further showed that participants in the hybrid mode more often selected exercises recommended by the adaptive mechanism rather than by GenAI. These findings suggest that although participants found GenAI-generated feedback helpful, they tended to prefer exercises recommended by the adaptive mechanism.

*6.2. Conclusion*

In conclusion, the findings of this research demonstrate that the hybrid GenAI–adaptive mode in ADVENTURE effectively supported learners in practicing programming and enhancing their programming skills. This mode offered flexibility by allowing learners to select the next exercise recommended either by the adaptive approach or the GenAI model, while also providing AI-generated feedback to support code comprehension. Additionally, the results showed that participants in the hybrid mode achieved the highest number of correct submissions, with most preferring to select exercises recommended by the adaptive approach. These findings highlight the value of combining adaptive mechanisms with GenAI feedback, offering a more effective approach to supporting programming education than either mode alone.

*6.3. Limitations*

This study has three primary limitations. The first concerns the exercise recommendation feature generated by the GenAI model. Although the prompt used in this study (as shown in Fig. A.1) instructed GPT-4 to avoid recommending exercises that appeared in previous conversations or chat history, some repetitions were still generated, which caused confusion for learners. This issue reflects a known limitation of LLMs, as they do not always accurately process the entire prompt input, particularly when it is lengthy or contains complex contextual information (Hosseini et al., 2024). While this study did not aim to resolve this limitation of LLMs, in order to mitigate learner confusion, participants were given the option to either repeat the exercise or proceed with a recommendation generated by the adaptive mechanism. This approach served as an alternative process in the GenAI mode and as the main process in the integration of adaptive and GenAI mode in this study. Future work will focus on optimizing prompt structure to reduce these inconsistencies in LLM-generated recommendations. The second limitation is the gender imbalance across the experimental groups, as mentioned in Section 4.1. As the objective of developing this system was to support novice learners in practicing programming skills, we recruited first- and second-year undergraduate students in Information Technology at a university in Thailand. However, this study faced a reduction in participant numbers, as only 123 (Male = 107 and Female = 16) of the 143 students who initially expressed interest to participate in the experiment, and ultimately only 87 (Male = 73 and Female = 14) completed both the experiment and the questionnaire. Given the characteristics of this target group, achieving balanced gender representation across all experimental groups was challenging. Therefore, this study focused on comparing experimental groups rather than analyzing gender differences within each group. However, we plan to expand the number of participants in future work to reach an improved group balance. Additionally, the third limitation concerns the translation of the questionnaire. The questionnaire used in this study was adapted from (Wang et al., 2014, 2020, 2024), originally developed in English, and was translated into Thai to conduct the experiment with undergraduate students at a university in Thailand. In future research, the Thai version will need to be reviewed to ensure the accuracy and clarity of the translated items and re-evaluated to guarantee the reliability and validity.

*6.4. Future work*

In terms of future work, to further enhance the effectiveness of ADVENTURE, several implementation strategies are proposed based on the findings of this study: (1) Expanding GenAI-generated feedback to include natural language explanations of code errors, along with an option to translate the feedback into other languages to enhance comprehension, and (2) the development of a learner-centered analytics dashboard to foster self-regulated learning by providing visual insights into individual progress, strengths, and recurring mistakes. Also, we plan to expand the participant pool by recruiting learners from multiple universities to create more balanced experimental groups. In addition, we aim to enhance some system features, such as preventing the recommendation of previously completed exercises and optimizing the difficulty level of exercises suggested by the GenAI model. Furthermore, we intend to explore strategies for leveraging GenAI to further improve learning outcomes and support skills development.

**CRediT authorship contribution statement**

**Lalita Na Nongkhai:** Writing – review & editing, Writing – original draft, Visualization, Validation, Software, Methodology, Investigation, Formal analysis, Data curation, Conceptualization. **Jingyun Wang:** Writing – review & editing, Visualization, Validation, Supervision, Project administration, Methodology, Formal analysis, Conceptualization. **Adam Wynn:** Writing – review & editing, Validation, Formal analysis. **Takahiko Mendori:** Writing – review & editing, Validation, Supervision, Resources, Project administration, Funding acquisition, Conceptualization.

**Ethics approval and consent to participate**

This study was approved by the Human Research Ethics Committee of the Kochi University of Technology and followed the relevant guidelines and regulations (protocol code: 283; date of approval: March 7, 2023).

**Materials and code availability**

The CONTINUOUS can be accessed from this website https://github.com/lalita-nk/CONTINUOUS.git.

**Ethics approval and consent to participate**

The study was approved by the Human Research Ethics Committee of the Kochi University of Technology and followed the relevant guidelines and regulations (protocol code: 283; date of approval: March 7, 2023). Informed consent was obtained from all participants, and their privacy rights were strictly observed. The data can be obtained by sending request e-mails to the corresponding author.

**Funding**

This research is funded by JSPS KAKENHI, grant number JP22K12318.

**Declaration of competing interest**

The authors declare that they have no known competing financial interests or personal relationships that could have appeared to influence the work reported in this paper.

**Acknowledgements**

We would like to acknowledge Mr. ANGGUO ZHOU for his support in the function of Generative AI integration, during his master project in Durham University.

**Appendix**

Figure A.1 shows the prompt used with GPT-4, which consists of five variables. These variables include:

(1) "question_content": the content of the current question.
(2) "correct_code": the answer provided by the instructor for the "question_content."
(3) "submitted_code": the code submitted by the learner.
(4) "chat_history": the previous responses generated by GPT-4 for the learner, and
(5) "context,": the background information for the next question recommended by GPT-4.

```
# main prompt
        combined_prompt = PromptTemplate(
            input_variables=["question_content", "correct_code", "submitted_code", "context", "chat_history"],
            template="""
            You are an instructor assisting a student in learning coding concepts.

            1. First, verify if the student's submitted code is correct based on the question_content and the correct_code provided:
               - question_content: {question_content}
               - Correct Code: {correct_code}
               - Submitted Code: {submitted_code}

            If the submitted code is incorrect:
               - Highlight the incorrect part of the code.
               - Show the correct code and explain the mistake. For example:

               Submitted Code:
               height = 7
               width = 5
               perimeter = 2 - (height + width) "The operation in the function is subtraction, not addition"
               print(perimeter)
               area = height * width
               print(area)

               the right part should be: perimeter = 2 * (height + width)

               Here is the correct code:
               height = 7
               width = 5
               perimeter = 2 * (height + width)
               print(perimeter)
               area = height * width

            2. Based on the previous conversation history:

               Chat History: {chat_history}

               explain why the student might have made this mistake and identify the knowledge points where the student may be struggling.

            3. Based on the student's performance, recommend a new exercise following the instructions below:

            - If the code is incorrect: Recommend exercises from the context provided below that address the specific mistake and expand the student's knowledge.
            - If the code is correct: Recommend more challenging exercises from the context provided below to further develop the student's understanding.

            Justify your recommendation by discussing the student's mastery of relevant knowledge points from previous conversations or Chat History.

            Warning:
            1.Do not recommend exercises that have appeared in previous conversations or Chat History.
            2.Do not include the 'code' content of the recommended exercise.

            Recommended Exercise:

            Question ID:
            Content:

            Recommended Reason:

            Note: The context provided below contains all exercises retrieved from the vector store to support recommendation.
            Context: {context}
            """
        )
```

**Fig. A.1.** Prompt that is used with GPT-4.

**Data availability**

The raw learner data generated during the current study are not publicly available due to individual user privacy concerns.